\documentclass[twocolumn,nofootinbib,amsmath,amssymb,a4paper,floatfix]{revtex4}
\usepackage{graphicx}
\usepackage{dcolumn}
\usepackage{bm}

\newcommand{\Ref}[1]{Ref.~\onlinecite{#1}}

\newcommand{\Fig}[1]{Fig.~\ref{#1}}
\newcommand{\Sec}[1]{Sec.~\ref{#1}}
\newcommand{\Secs}[1]{Secs.~\ref{#1}}
\newcommand{\Tab}[1]{Table~\ref{#1}}
\newcommand{\Eq}[1]{Eq.~(\ref{#1})}

\newcommand{\eg}{{\it e.g.}}

\newcommand{\BR}[1]{\ensuremath{{\rm BR}(#1)}}

\newcommand{\order}[1]{\ensuremath{\mathcal{O}(#1)}}
\newcommand{\SN}[2]{\ensuremath{#1\times10^{#2}}}

\newcommand{\Vud}{\ensuremath{|V_{ud}|}}
\newcommand{\Vus}{\ensuremath{|V_{us}|}}
\newcommand{\Vusf}{\ensuremath{|V_{us}|f_+(0)}}

\begin{document}

\title{$V_{us}$ from kaon decays}

\author{Matthew Moulson}
\email{Matthew.Moulson@lnf.infn.it}
\affiliation{Laboratori Nazionali di Frascati dell'INFN\\
00044 Frascati (Roma), Italy}
\author{for the FlaviaNet Working Group on Kaon Decays}
\homepage{http://www.lnf.infn.it/wg/vus}
\noaffiliation

\begin{abstract}
Recent measurements of kaon decays contributing to the determination of 
\Vus\ are summarized, and up-to-date evaluations of \Vusf\ and \Vus\
are presented.
\end{abstract}

\maketitle

\section{Introduction}
At present, the first-row constraint 
$|V_{ud}|^2 + |V_{us}|^2 + |V_{ub}|^2$ (with $|V_{ub}|^2$ negligible)
offers the most precise test of CKM unitarity.
Up until 2002 (and for the 2004 PDG evaluation~\cite{PDG04:Vus}), 
the evaluation
of \Vus\ from older $K_{\ell3}$ data gave a $2.3\sigma$ hint of unitarity 
violation in the first-row test. The 2003 measurement of 
\BR{K^+_{e3}} by BNL E865~\cite{E865+03:Ke3} gave a value for \Vus\ 
consistent with 
unitarity. In the period 2004--2006, many new measurements of BRs, 
lifetimes, and form-factor slopes were announced by KLOE, KTeV, 
ISTRA+, and NA48. All of these new measurements are distinguished from
the older measurements in that they are based on much higher statistics, 
and in that radiative corrections are applied consistently.
Late-2005 evaluations of \Vus\ by the CKM working group on the first 
row~\cite{CKM05} and in the 2006 PDG review~\cite{PDG06:Vus} include many, 
but not all, of these important developments. We present an
up-to-date evaluation that includes preliminary results presented at 
this conference.

\Vus\ is evaluated from $K_{\ell3}$ data using the relation
\begin{eqnarray}
\Gamma(K_{\ell3(\gamma)})& =& 
\frac{C_K^2G_F^2M_K^5}{192\pi^3}\,S_{\rm EW}\,\Vus^2\,|f_+(0)|^2 \times
\nonumber \\
& & I_{K\ell}({\lambda}_{K\ell})\:(1 + 2\Delta_K^{SU(2)} + 2\Delta_{K\ell}^{\rm EM}),
\label{eq:Vus}
\end{eqnarray}
where $C_K$ is a Clebsch-Gordan coefficient; $M_K$ is the appropriate kaon 
mass; $S_{\rm EW} = 1.0232$ is the universal short-distance electroweak 
correction; $f_+(0) \equiv f_+^{K^0\pi^-}(0)$ is the 
value of the hadronic matrix element (form factor) for the $K^0\to\pi^-$ 
transition at zero 
momentum transfer to the leptonic system; $I_{K\ell}$ is the phase-space 
integral of the normalized form factor, which depends on the values of 
one or more slope parameters $\lambda$; $\Delta_K^{SU(2)}$ 
and $\Delta_{K\ell}^{\rm EM}$ are $SU(2)$-breaking and 
long-distance electromagnetic corrections; and the subscripts 
$K$ and $\ell$ indicate dependence on the kaon charge and lepton flavor.
Our main result is a current average value for \Vusf. We first review 
the recent data on $K_{\ell3}$ rates (BRs and lifetimes) and form-factor 
slopes.
 
\section{$K_L$ decay rate data}
\label{sec:KL}

Numerous measurements of the principal $K_L$ BRs, or of various ratios
of these BRs, have been published recently. For the purposes of evaluating
\Vusf, these data can be used in a PDG-like fit to the $K_L$ BRs and lifetime,
so all such measurements are interesting.
   
KTeV has measured five ratios of the six main $K_L$ BRs~\cite{KTeV+04:BR}. 
The six channels
involved account for more than 99.9\% of the $K_L$ width and KTeV combines the
five measured ratios to extract the six BRs. We use the five measured ratios
in our analysis: 
$\BR{K_{\mu3}/K_{e3}} = 0.6640(26)$,  
$\BR{\pi^+\pi^-\pi^0/K_{e3}} = 0.3078(18)$,  
$\BR{\pi^+\pi^-/K_{e3}} = 0.004856(28)$,  
$\BR{3\pi^0/K_{e3}} = 0.4782(55)$, and
$\BR{2\pi^0/3\pi^0} = 0.004446(25)$. The errors on these measurements are
correlated; this is taken into account in our fit.  

NA48 has measured the ratio of the BR for $K_{e3}$ decays to the sum of BRs
for all decays to two tracks, giving 
$\BR{K_{e3}}/(1-\BR{3\pi^0}) = 0.4978(35)$ \cite{NA48+04:BR}. From a 
separate measurement of \BR{K_L\to3\pi^0}/\BR{K_S\to2\pi^0}, NA48 
obtains $\BR{3\pi^0}/\tau_L = 3.795(58)$~MHz \cite{Lit04:ICHEP}.  

Using $\phi\to K_L K_S$ decays in which the $K_S$ decays to $\pi^+\pi^-$,
providing normalization, KLOE has directly measured the BRs for the four 
main $K_L$ decay channels \cite{KLOE+06:BR}.
The errors on the KLOE BR values are dominated 
by the uncertainty on the $K_L$ lifetime $\tau_L$; since the dependence of 
the geometrical efficiency on $\tau_L$ is known, KLOE can solve for $\tau_L$
by imposing $\sum_x \BR{K_L\to x} = 1$ (using previous averages for the minor 
BRs), thereby greatly reducing the uncertainties on the BR values obtained.
Our fit makes use of the KLOE BR values before application of this constraint:
\BR{K_{e3}} = 0.4049(21),   
\BR{K_{\mu3}} = 0.2726(16),   
\BR{K_{e3}} = 0.2018(24), and
\BR{K_{e3}} = 0.1276(15).
The dependence of these values on $\tau_L$ and the correlations between the 
errors \cite{Spa06:CKM} are taken into account.

KLOE has also measured $\tau_L$ directly, by fitting the proper decay time
distribution for $K_L\to3\pi^0$ events, for which the reconstruction 
efficiency is high and uniform over a fiducial volume of $\sim$$0.4\lambda_L$.
They obtain $\tau_L=50.92(30)$~ns \cite{KLOE+05:tauL}. 

There are also two recent measurements of \BR{\pi^+\pi^-/K_{\ell3}}, 
in addition to the KTeV measurement of \BR{\pi^+\pi^-/K_{e3}} discussed above.
KLOE obtains \BR{\pi^+\pi^-/K_{\mu3}} = \SN{7.275(68)}{-3} \cite{KLOE+06:KLpp},
while NA48 obtains \BR{\pi^+\pi^-/K_{e3}} = \SN{4.826(27)}{-3} 
\cite{NA48+06:KLpp}. All measurements are fully inclusive of inner 
bremsstrahlung. The KLOE measurement is fully inclusive of the direct-emission 
(DE) component, DE contributes negligibly to the KTeV measurement, and a 
residual DE contribution of 0.19\% has been subtracted from the NA48 value 
to obtain the number quoted above. For consistency, in our fit, 
a DE contribution of 1.52(7)\% is added to the KTeV and NA48 values.
Our fit result for \BR{\pi^+\pi^-} is then understood to be DE inclusive. 

In addition to the 14 recent measurements listed above, our fit for the 
seven largest $K_L$ BRs and lifetime uses four of the remaining five 
inputs to the 2006 PDG fit and the constraint that the seven BRs sum to unity. 
The results are summarized in \Tab{tab:KLBR}.
\begin{table}
\caption{\label{tab:KLBR}
Results of fit to $K_L$ BRs and lifetime}
\begin{ruledtabular}
\begin{tabular}{lcr}
Parameter & Value & $S$ \\
\hline
\BR{K_{e3}} & 0.40563(74) & 1.1 \\
\BR{K_{\mu3}} & 0.27047(71) & 1.1 \\
\BR{3\pi^0} & 0.19507(86) & 1.2 \\
\BR{\pi^+\pi^-\pi^0} & 0.12542(57) & 1.1 \\
\BR{\pi^+\pi^-} & \SN{1.9966(67)}{-3} & 1.1 \\
\BR{2\pi^0} & \SN{8.644(42)}{-4} & 1.3 \\
\BR{\gamma\gamma} & \SN{5.470(40)}{-4} & 1.1 \\
$\tau_L$ & 51.173(200)~ns & 1.1 \\
\end{tabular}
\end{ruledtabular}
\end{table}

Our fit gives $\chi^2/{\rm ndf} = 20.2/11$ (4.3\%), while the 2006 
PDG fit gives $\chi^2/{\rm ndf} = 14.8/10$ (14.0\%).
The differences between the output values from our fit and the 
2006 PDG fit are minor.
The poorer value of $\chi^2/{\rm ndf}$ for our fit can be traced 
to contrast between the KLOE value for \BR{3\pi^0} and the 
other inputs involving \BR{3\pi^0} and \BR{\pi^0\pi^0}---in 
particular, the PDG ETAFIT value for \BR{\pi^0\pi^0/\pi^+\pi^-}.
The treatment of the correlated KTeV and KLOE measurements in the 
2006 PDG fit gives rise to large 
scale factors for \BR{K_{e3}} and \BR{3\pi^0}; 
in our fit, the scale factors are more uniform. As a result, 
our value for \BR{K_{e3}} has a significantly smaller uncertainty
than does the 2006 PDG value.

\section{$K_S$ decay rate data}
\label{sec:KS}

Two KLOE measurements provide the BR for the $K_{e3}$ decay of the $K_S$
with enough precision to be of interest for the determination of \Vus.
Using $\phi\to K_L K_S$ decays in which the $K_L$ is recognized by its
interaction in the experiment's EM calorimeter, KLOE measures 
\BR{K_{e3}/\pi^+\pi^-} = \SN{10.19(13)}{-4} \cite{KLOE+06:KSe3}.
KLOE has also recently obtained \BR{\pi^+\pi^-/\pi^0\pi^0} = 2.2459(54)
\cite{KLOE+06:Kspp}, 
where this value is an average including the experiment's 2002 result.
These measurements completely determine the main $K_S$ BRs, and
in particular give $\BR{K_{e3}} = \SN{7.046(91)}{-4}$ \cite{KLOE+06:KSe3}.
Our evaluation of \Vusf\ uses this value, together with the lifetime value 
$\tau_S = 0.08958(5)$~ns from the PDG fit to $CP$ parameters, which is
highly constrained by measurements from NA48 \cite{NA48+02:tauS} and 
KTeV \cite{KTeV+03:CPpar}.  

\section{$K^\pm$ decay rate data}
\label{sec:Kpm}

There are several new results providing information on $K^\pm_{\ell3}$ 
rates. These results are mainly preliminary and have not been included 
in previous averages.

NA48/2 has recently submitted for publication measurements of the three 
ratios
\BR{K_{e3}/\pi\pi^0},  
\BR{K_{\mu3}/\pi\pi^0}, and  
\BR{K_{\mu3}/K_{e3}} \cite{Vel06:CKM,NA48+07:BR}. 
These measurements are not independent; in our fit, we use the values 
$\BR{K_{e3}/\pi\pi^0} = 0.2496(10)$ and   
$\BR{K_{\mu3}/\pi\pi^0} = 0.1637(7)$ and take their correlation
into account.

ISTRA+ has also updated its preliminary value for $\BR{K_{e3}/\pi\pi^0}$.
They now quote $\BR{K_{e3}/\pi\pi^0} = 0.2449(16)$, as reported
at this conference \cite{Rom06:CKM}.

KLOE has measured the absolute BRs for the
$K_{e3}$ and $K_{\mu3}$ decays
\cite{Sci06:Lisbon}.
In $\phi\to K^+ K^-$ events, $K^+$ decays into $\mu\nu$ or $\pi\pi^0$
are used to tag a $K^-$ beam, and vice versa. KLOE performs four 
separate measurements for each $K_{\ell3}$ BR, corresponding to the
different combinations of kaon charge and tagging decay.
The final averages are $\BR{K_{e3}} = 5.047(92)\%$ and
$\BR{K_{\mu3}} = 3.310(81)\%$.
Our fit takes into
account the correlation between these values, as well as their dependence
on the $K^\pm$ lifetime \cite{Spa06:CKM}.

The world average value for $\tau_\pm$ is nominally 
quite precise; the 2006 PDG quotes $\tau_\pm = 12.385(25)$~ns.
However, the error is scaled by 2.1; the confidence level for the 
average is 0.2\%. It is important to confirm the value of $\tau_\pm$.
A preliminary measurement from KLOE,
$\tau_\pm = 12.367(78)$~ns \cite{Spa06:CKM}, agrees with the PDG 
average, although at present the KLOE uncertainty is significantly larger.

Our fit for the six largest $K^\pm$ BRs and lifetime makes use of the 
results cited above, 
plus the data used in the 2006 PDG fit, for a total of 30 measurements.
The six BRs are constrained to sum to unity.
The results are summarized in \Tab{tab:KpmBR}.
\begin{table}
\caption{\label{tab:KpmBR}
Results of fit to $K^\pm$ BRs and lifetime}
\begin{ruledtabular}
\begin{tabular}{lcr}
Parameter & Value & $S$ \\
\hline
\BR{K_{\mu2}}      & 63.442(145)\%   & 1.3 \\
\BR{\pi\pi^0}      & 20.701(108)\%   & 1.3 \\
\BR{\pi\pi\pi}     &  5.5921(305)\%  & 1.0 \\
\BR{K_{e3}}        &  5.121(38)\%    & 1.6 \\
\BR{K_{\mu3}}      &  3.3855(203)\%  & 1.2 \\
\BR{\pi\pi^0\pi^0} &  1.7592(234)\%  & 1.1 \\
$\tau_\pm$         & 12.3840(213)~ns & 1.8 \\
\end{tabular}
\end{ruledtabular}
\end{table}
\begin{figure*}
\includegraphics[width=0.8\textwidth]{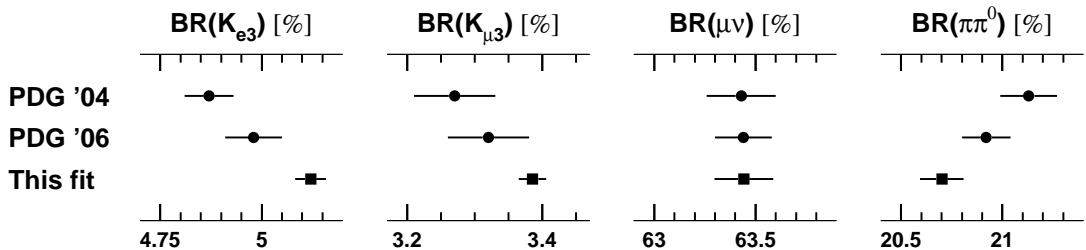}
\caption{\label{fig:kpmavg}
Evolution of average values for main $K^\pm$ BRs.}
\end{figure*}

The fit quality is poor, with $\chi^2/{\rm ndf} = 49/24$ (0.21\%).
However, when the five older measurements of $\tau_\pm$ are replaced 
by their PDG average with scaled error,  $\chi^2/{\rm ndf}$ improves 
to 31.3/20 (5.1\%), with no significant changes in the results.
Tension between the new measurements involving \BR{K_{e3}} and 
the older measurements is partly responsible for the poor fit quality
(note the scale factor of 1.6 on \BR{K_{e3}} from our fit).
Because of uncertainties concerning the treatment of radiative corrections,
it would be desirable to eliminate some of the older measurements, 
as both we and PDG do for the $K_L$ fit. However, for $K^\pm$,
all of the new measurements involve the $K_{\ell3}$ BRs or the 
ratios \BR{K_{\ell3}/\pi\pi^0}. 
This leads to large correlations
between the $K_{\ell3}$ BRs and \BR{\pi\pi^0}, and makes the fit
unstable when the older data are excluded.
Both the significant evolution of the average values of the $K_{\ell3}$ 
BRs and the effect of the correlations with \BR{\pi\pi^0} are evident
in \Fig{fig:kpmavg}. 

\section{Form-factor slopes}
\label{sec:FF}

For $K_{e3}$ decays, recent measurements of the quadratic slope parameters 
of the vector form factor $({\lambda_+',\lambda_+''})$ are available from 
KTeV \cite{KTeV+04:FF,KTeV+06:Hill}, 
KLOE \cite{KLOE+06:FF}, ISTRA+ \cite{ISTRA+04:e3FF,Rom06:CKM}, and 
NA48 \cite{NA48+04:e3FF}.
For $K_{\mu3}$ decays, results of fits using the quadratic parameterization
for the vector form factor $({\lambda_+',\lambda_+''})$ and the linear 
parameterization for the scalar form factor $(\lambda_0)$ are available
from 
KTeV \cite{KTeV+04:FF}, 
ISTRA+ \cite{ISTRA+04:m3FF,Rom06:CKM}, and 
NA48 \cite{Vel06:CKM,NA48+07:m3FF}.
ISTRA+ measures $K^-$ decays; the other experiments measure $K_L$ decays.
\begin{figure}
\includegraphics[width=0.4\textwidth]{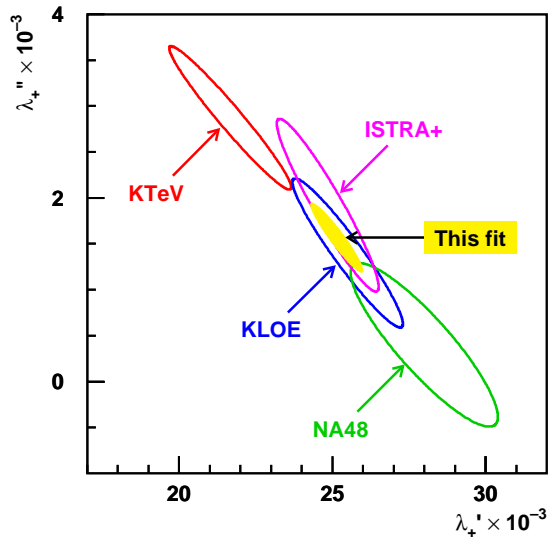}
\caption{\label{fig:ke3ff}
Recent measurements of form-factor slopes for $K_{e3}$ decays, 
with fit.}
\end{figure}
\begin{figure*}
\includegraphics[width=0.85\textwidth]{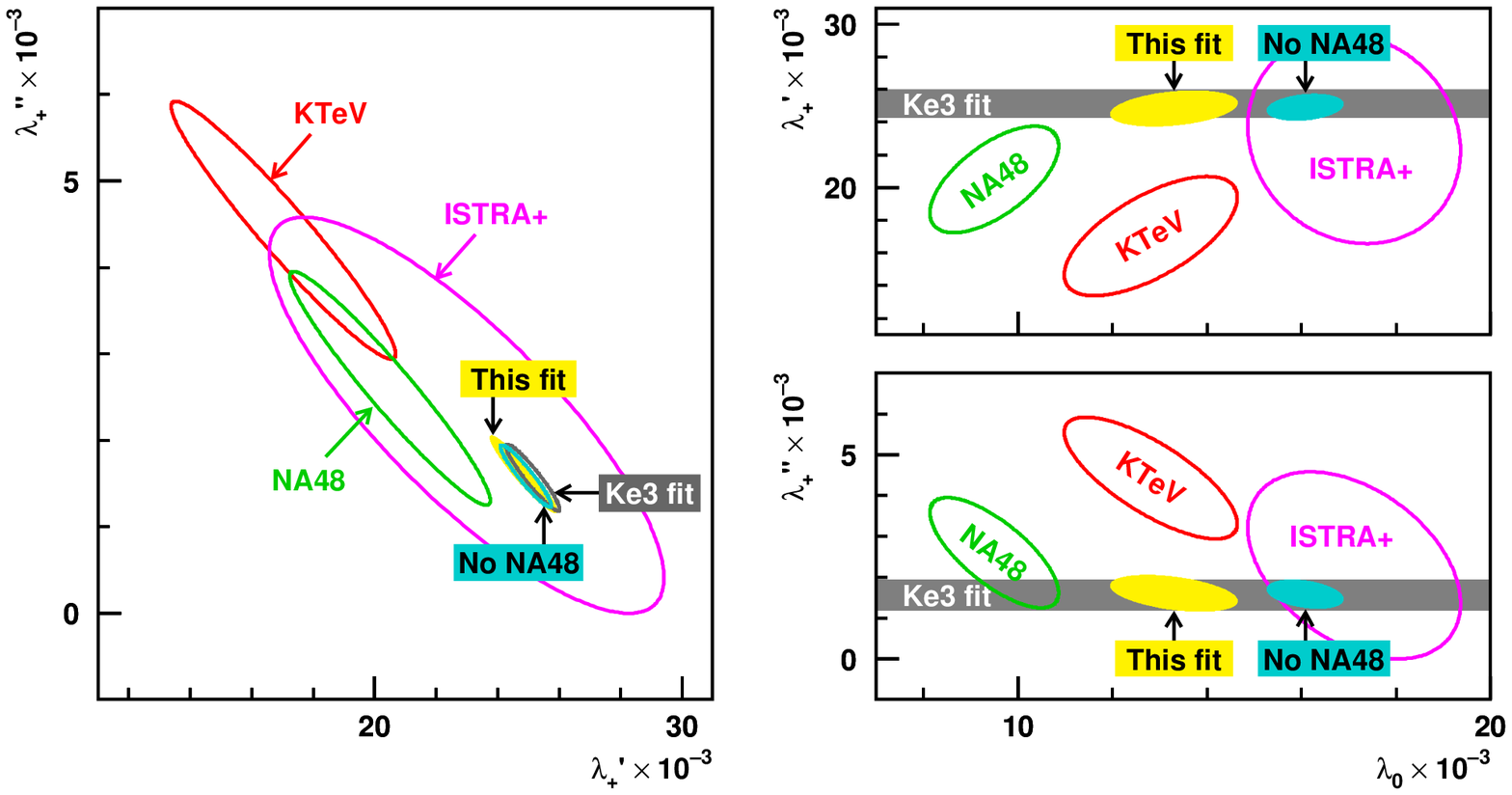}
\caption{\label{fig:kl3ff}
Recent measurements of form-factor slopes for $K_{\ell3}$ decays, 
with fits.}
\end{figure*}
 
The $K_{e3}$ data are summarized in \Fig{fig:ke3ff}. 
The uncertainties on the values of $\lambda_+'$ and $\lambda_+''$ reported
by each experiment are highly correlated. 
This is an intrinsic property of the quadratic parameterization. 
The form-factor slopes represent a small modification of
the kinematic density of the matrix element for $K\to\pi$ transitions;
in addition, sensitivity to $\lambda_+''$ is poor because the kinematic
density of the matrix element drops to zero at large $t$, where the form
factor itself is maximal. Taking the correlations into account, 
we obtain a good fit to the data in \Fig{fig:ke3ff}:
$\lambda_+' = \SN{24.15(0.87)}{-3}$ and
$\lambda_+'' = \SN{1.57(0.38)}{-3}$, with
$\rho(\lambda_+',\lambda_+'') = -0.941$ and 
$\chi^2/{\rm ndf} = 5.3/6$ (51\%). 
The significance of the quadratic term is greater than $4\sigma$.
The fit result is shown as the yellow ellipse in \Fig{fig:ke3ff}. 

The $K_{\mu3}$ data are summarized in \Fig{fig:kl3ff}, which also shows
the results for $\lambda_+'$ and $\lambda_+''$ from our $K_{e3}$ fit, and
the results of our combined fits to $K_{e3}$ and $K_{\mu3}$ data.
While the results for $\lambda_+'$ and $\lambda_+''$ from $K_{\mu3}$
decays are fairly consistent and reasonably agree with those from 
$K_{e3}$ decays (\Fig{fig:kl3ff}, left panel), 
the results for $\lambda_0$ show poor 
consistency (right panels). The most discrepant measurement is that of 
NA48. In both the $\lambda_+'$-$\lambda_0$ and $\lambda_+''$-$\lambda_0$
views, a locus of values within the $1\sigma$ confidence contours from ISTRA+
and our $K_{e3}$ fit lies along the major axis of the KTeV confidence 
ellipse. As a result, a fit to all data ($K_{e3}$ and $K_{\mu3}$)
excluding only 
the $K_{\mu3}$ results from NA48 gives $\chi^2/{\rm ndf} = 11.9/9$ (21.7\%).
The results of this fit are indicated by the cyan ellipses in \Fig{fig:kl3ff}. 
When the NA48 $K_{\mu3}$ results are included,
$\chi^2/{\rm ndf}$ increases to $52/12$, for a probability of less than 
$10^{-6}$. The results are shown as the yellow ellipses in \Fig{fig:kl3ff}.

Since at the moment there is no a priori reason to exclude the NA48 $K_{\mu3}$ 
results, we base our estimate of \Vusf\ on the fit to all data. However, we
scale the errors on the slope parameters to reflect the inconsistency in
the input data set. We obtain 
$\lambda_+' = \SN{24.84(1.10)}{-3}$ $(S=1.4)$,
$\lambda_+'' = \SN{1.61(0.45)}{-3}$ $(S=1.3)$, and
$\lambda_0 = \SN{13.30(1.35)}{-3}$ $(S=2.1)$, with
$\rho(\lambda_+',\lambda_+'') = -0.944$,
$\rho(\lambda_+',\lambda_0) = +0.314$, and 
$\rho(\lambda_+'',\lambda_0) = -0.420$.
From these results, we calculate the phase-space integrals for use 
with \Eq{eq:Vus} to be
$I(K^0_{e3}) = 0.15452(29)$,
$I(K^\pm_{e3}) = 0.15887(30)$,
$I(K^0_{\mu3}) = 0.10207(34)$, and
$I(K^\pm_{\mu3}) = 0.10501(35)$.

\section{Evaluation of \Vusf}

The strong $SU(2)$-breaking and EM corrections to $f_+(0)$ 
used in our evaluation of \Vusf\ are summarized in 
\Tab{tab:corr}.
\begin{table}
\caption{\label{tab:corr}
Summary of $SU(2)$-breaking and EM corrections}
\begin{ruledtabular}
\begin{tabular}{lcc}
Mode           & $\Delta^{SU(2)}$ & $\Delta^{\rm EM}$ \\ \hline
$K^0_{e3}$     & 0                & $+0.52(10)\%$ \\
$K^\pm_{e3}$   & $+2.31(22)\%$    & $+0.03(10)\%$ \\
$K^0_{\mu3}$   & 0                & $+0.95(15)\%$ \\
$K^\pm_{\mu3}$ & $+2.31(22)\%$    & $+0.2(4)\%$ \\
\end{tabular}
\end{ruledtabular}
\end{table}
For $\Delta^{SU(2)}$, we use the standard value from \Ref{C+02:Kl3rad}.
All of the recent $K_{\ell3}$ BR measurements are fully inclusive 
of final-state radiation; the values we use for $\Delta_{K\ell}^{\rm EM}$
were calculated for the fully-inclusive rates.
Specifically,  for $K^0_{e3}$ and $K^\pm_{e3}$, the values were obtained 
from the chiral perturbation theory calculation of \Ref{C+02:Kl3rad}, 
updated using the evaluations of the low-energy constants 
from \Ref{DGM05:LEC}.
For $K^0_{\mu3}$, the value is from \Ref{And04:KLOR}, and was obtained 
using a generator implementing a hadronic-model calculation.
For $K^\pm_{\mu3}$, we do not know of any complete estimate.
Our value is loosely based on \Ref{Gin70:rad}, with a very generous
error estimate.
However, with this treatment, $\Delta_{K\ell}^{\rm EM}$ gives the 
largest single contribution to the uncertainty on \Vusf\ from this mode.

We use the results of the fits and averages described in 
\Secs{sec:KL}--\ref{sec:FF}, together with the corrections in 
\Tab{tab:corr}, to evaluate the quantity \Vusf\ for each of the 
five $K_{\ell3}$ decay modes as listed in \Tab{tab:Vusf}.
\begin{table}
\caption{\label{tab:Vusf}
Values of \Vusf\ from data for different decay modes, with error breakdown}
\begin{ruledtabular}
\begin{tabular}{lcccccc}
              &              &         & \multicolumn{4}{@{}c@{}}{\scriptsize Approx contrib to \% err} \\
Mode          & \Vusf\       & \% err  & BR   & $\tau$ & $\Delta$ & $I$  \\ 
\hline
$K_L\:e3$     & 0.21639(55)  & 0.25    & 0.09 & 0.19   & 0.10     & 0.09 \\ 
$K_L\:\mu3$   & 0.21649(68)  & 0.31    & 0.10 & 0.18   & 0.15     & 0.17 \\
$K_S\:e3$     & 0.21555(142) & 0.66    & 0.65 & 0.03   & 0.10     & 0.09 \\
$K^\pm\:e3$   & 0.21844(101) & 0.46    & 0.38 & 0.11   & 0.24     & 0.09 \\
$K^\pm\:\mu3$ & 0.21809(125) & 0.57    & 0.31 & 0.10   & 0.45     & 0.17 \\
\end{tabular}
\end{ruledtabular}
\end{table}
The table also gives the fractional uncertainty for each determination
of \Vusf, as well as the approximate breakdown of the contributions
to the uncertainty from the various inputs (these contributions are 
understood to be added in quadrature). The source of the limiting 
uncertainty varies from mode to mode. 

The average value of \Vusf\ from all modes, as obtained from a fit 
with correlations taken into account, is 0.21673(46).
The fit gives $\chi^2/{\rm ndf} = 4.2/4$ (38\%).

Although the value of $\chi^2/{\rm ndf}$ from the fit is satisfactory, 
one notes that the values of \Vusf\ for the two charged modes seem to be 
higher than those for the neutral modes. To quantify this, we perform the 
fit separately for charged and netural modes, using the results of the 
overall fit to the form-factor slope data of \Sec{sec:FF}, and the 
$SU(2)$-breaking correction from \Tab{tab:corr}, 
$\Delta^{SU(2)}_{\rm th}=2.31(22)\%$, for the charged modes.
We obtain $\Vusf = 0.21635(50)$ for neutral modes and 0.21832(94) for 
charged modes, a $1.9\sigma$ difference. If we take this as a suggestion
that the $SU(2)$-breaking correction may be underestimated, and perform
the fit leaving free the value of \Vusf\ for neutral modes and an 
empirical value of $\Delta^{SU(2)}$, we obtain
$\Delta^{SU(2)}_{\rm exp}=3.24(43)\%$.

Comparison of the values of \Vusf\ for $K_{e3}$ and $K_{\mu3}$ modes 
provides a test of lepton universality. Specifically, 
\begin{displaymath}
r_{\mu e} 
= \frac{[|V_{us}|\,f_+(0)]^2_{\mu3,\:{\rm exp}}}
       {[|V_{us}|\,f_+(0)]^2_{e3,\:{\rm exp}}} 
= \frac{\Gamma_{\mu3}}{\Gamma_{e3}}\cdot
  \frac{I_{e3}\,(1+\delta_{e3})}{I_{\mu3}\,(1+\delta_{\mu3})}. 
\end{displaymath}
By comparison with \Eq{eq:Vus}, $r_{\mu e}$ is equal to the ratio
$g_\mu^2/g_e^2$, with $g_\ell$ the coupling strength at the 
$W\to\ell\nu$ vertex. In the Standard Model, $r_{\mu e} = 1$.
We calculate $r_{\mu e}$ separately for charged and neutral modes.
The results are compatible; the average value is $r_{\mu e} = 1.0003(53)$.
We note that the sensitivity of this test of lepton universality
is beginning to approach that obtained in $\pi\to\ell\nu$ decays:
$(r_{e \mu})_{\pi_{\ell2}} = 0.9966(30)$, as reviewed in \Ref{ERM05:lowE}.

\section{Evaluation of \Vus}

To test CKM unitarity, a value for $f_+(0)$ is needed.
In the chiral expansion, $f_+(0) = 1 + f_{p^4} + f_{p^6} + \cdots$.
A finite one-loop contribution gives rise to $f_{p^4}$; its calculation
\cite{LR84:f0,GL85:f0} involves essentially no uncertainty.
A quark-model estimate of the contribution at \order{p^6}
was originally performed by Leutwyler and Roos \cite{LR84:f0}.
More recently, a complete \order{p^6} calculation in chiral perturbation
theory showed that meson-loop contributions are sizeable \cite{BT03:f0}. 
Analytic estimates of $f_+(0)$ were surveyed 
at this conference \cite{Por06:CKM}, and
lattice evaluations of $f_+(0)$ are rapidly improving
in precision (see, \eg, \Ref{Zan06:CKM}).
Until a definite consensus emerges on a new reference value for $f_+(0)$,
however,
we use the original estimate of Leutwyler and Roos,
$f_+(0) = 0.961(8)$, which is supported by lattice calculations 
\cite{Zan06:CKM}. This gives $\Vus=0.2255(19)$.
\begin{figure}
\includegraphics[width=0.4\textwidth]{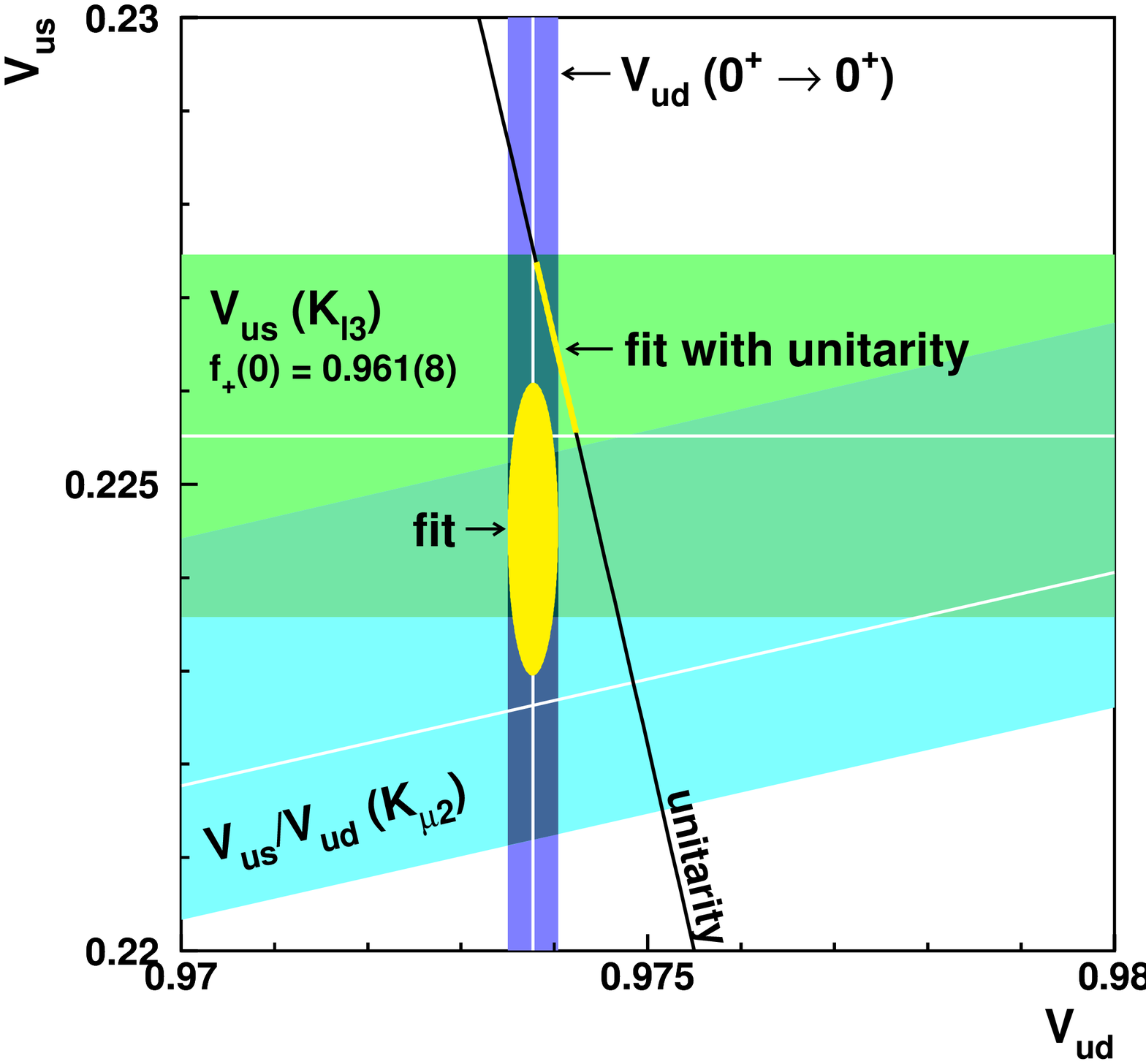}
\caption{\label{fig:univers}
Results of fits to \Vud, \Vus, and $\Vus/\Vud$.}
\end{figure}

Using the recent evaluation of \Vud\ from $0^+\to0^+$ 
nuclear beta decays (\Ref{MS06:Vud}; see also \Ref{Har06:CKM}),
$\Vud = 0.97377(27)$,
one has $\Vud^2 + \Vus^2 - 1 = -0.0009(10)$, a result
compatible with unitarity.

Marciano \cite{Mar04:fKpi} has observed that 
$\Gamma(K_{\mu2})/\Gamma(\pi_{\mu2})$ can be precisely related to 
the product $(\Vus/\Vud)^2(f_K/f_\pi)^2$. The recent
measurement $\BR{K^+\to\mu^+\nu}=0.6366(17)$ from KLOE \cite{KLOE+06:Km2},
together with the preliminary lattice result 
$f_K/f_\pi = 1.208(2)(^{+7}_{-14})$ from the MILC Collaboration 
\cite{Ber06:Chiral}, gives $\Vus/\Vud = 0.2286(^{+27}_{-15})$.
This ratio can be used in a fit together with the values of \Vud\
from \Ref{MS06:Vud} and \Vus\ from $K_{\ell3}$ decays as above.
Using the value for \Vus\ obtained with $f_+(0)=0.961(8)$, 
the fit gives $\Vud = 0.97377(27)$ and $\Vus = 0.2245(16)$, 
with $\chi^2/{\rm ndf} = 0.77/1$ (38\%). 
The unitarity constraint can also be included, in which 
case the fit gives $\chi^2/{\rm ndf} = 3.10/2$ (21.2\%, or $1.2\sigma$).
Both results are illustrated in \Fig{fig:univers}.

In summary, from $K_{\ell3}$ data we obtain $\Vusf = 0.21673(46)$,
where the uncertainty amounts to 0.21\%.
The dominant contribution to the uncertainty on \Vus\ is from $f_+(0)$.
Whether \Vus\ is evaluated from $K_{\ell3}$ data alone or with the 
additional constraint from $K_{\mu2}$ decays, the first-row unitarity
test is satisfied at about the $1\sigma$ level. 

\begin{acknowledgments}
FlaviaNet is a Marie Curie Research Training Network supported by the 
European Union Sixth Framework Programme under contract MTRN-CT-2006-035482.
\end{acknowledgments}

\bibliography{moulson}

\end{document}